\documentclass[a4paper]{PoS}

\title{The sub-TeV transient Gamma-Ray sky: challenges and opportunities}

\ShortTitle{Sub-TeV transient Gamma-Ray sky}

\author{Giovanni La Mura\thanks{Contact author.},$^a$ Pedro Assis,$^b$ Alberto Blanco,$^c$ \speaker{Ruben Concei\c{c}\~ao},$^b$ Paulo Fonte,$^c$ Lu\'{i}s Lopes,$^c$ M\'ario Pimenta,$^b$ Bernardo Tom\'e,$^b$ Catarina Esp\'{i}rito Santo,$^a$ Lu\'{i}s Mendes,$^a$ Miguel Ferreira,$^a$ Pedro Abreu,$^b$ Pedro Brogueira,$^d$ Lu\'{i}s Filipe Mendes,$^d$ Fernando Bar\~ao,$^b$ Ulisses Barres de Almeida,$^e$ Ronald Cintra Shellard,$^e$ Ugo Giaccari,$^e$ Otto Lippmann,$^e$ Benedetto D'Ettorre Piazzoli,$^f$ Michele Doro,$^g$ Elisa Prandini,$^g$ C\'edric Perennes,$^g$ Giorgio Matthiae,$^h$ Marco Tavani,$^i$ Rinaldo Santonico,$^j$ Alessandro De Angelis,$^k$ Ruben Lopez Coto,$^l$ Andrea Chiavassa,$^m$ Jakub V\'{i}cha,$^n$ Petr Tr\'avni\v{c}\'ek,$^n$ Giuseppe Di Sciascio$^o$ \\
  \llap{$^a$} LIP Lisbon, Av. Prof. Gama Pinto 2, 1649-003 Lisboa, Portugal\\
  \llap{$^b$} LIP/IST Lisbon, Av. Prof. Gama Pinto 2, 1649-003 Lisboa, Portugal\\
  \llap{$^c$} LIP Coimbra, Rua Larga 3004-516 Coimbra, Portugal\\
  \llap{$^d$} IST Lisbon, Av. Rovisco Pais 1, 1049-001 Lisboa, Portugal\\
  \llap{$^e$} CBPF, Rua Dr. Xavier Siguad, 150 - Urca - Rio de Janeiro - RJ, Brazil\\
  \llap{$^f$} Universit\`a di Napoli ``Federico II'' / INFN, Corso Umberto I 40, 80138 Napoli, Italy\\
  \llap{$^g$} University and INFN Padova, via Marzolo 8, I-35131 Padova, Italy\\
  \llap{$^h$} Universit\`a di Roma ``Tor Vergata'' / INFN, Via Cracovia 120, 00133 Roma, Italy\\
  \llap{$^i$} IAPS-INAF, Via Fosso del Cavaliere 100, 00133 Roma, Italy\\
  \llap{$^j$} INFN, Piazzale Aldo Moro 2, 00185 Roma, Italy\\
  \llap{$^k$} Universit\`a degli Studi di Udine / INFN, Via Palladio 8, 33100 Udine, Italy\\
  \llap{$^l$} INFN Padova, Via Marzolo 8, 35131 Padova, Italy\\
  \llap{$^m$} Universit\`a degli Studi di Torino / INFN, Via Verdi 8, 10124 Torino, Italy\\
  \llap{$^n$} Institute of Physics of the Czech Academy of Sciences, Na Slovance 1999/2, 182 21 Prague 8, Czech Republic\\
  \llap{$^o$} INFN Roma Tor Vergata, Via della Ricerca Scientifica 1, 00133 Rome, Italy\\
        E-mail: \email{glamura@lip.pt}}

\abstract{The detection of gravitational waves and neutrinos from astrophysical sources with gamma-ray counterparts officially started the era of Multi-Messenger Astronomy. Their transient and extreme nature implies that monitoring the VHE sky is fundamental to investigate the non-electromagnetic signals. However, the limited effective area of space-borne instruments prevents observations above a few hundred GeV, while the small field of view and low duty cycle of IACTs make them unsuited for extensive monitoring activities and prompt response to transients. Extensive Air Shower arrays (EAS) can provide a large field of view, a wide effective area and a very high duty cycle. Their main difficulty is the distinction between gamma-ray and cosmic-ray initiated air showers, especially below the TeV range.

  Here we present some case studies stressing the importance that a new EAS array in the Southern Hemisphere will be able to survey the sky from below 100 GeV up to several TeV. In the energy domain between 100 and 400 GeV we expect the strongest electromagnetic signatures of the acceleration of ultra-relativistic particles in sources like SNRs, blazar jets and gamma-ray bursts, as recently proved by IACT observations. This spectral window is also crucial to understand the Universe opacity to high energy radiation, thus providing constraints on the cosmological parameters. We will discuss the implications of VHE radiation on the mechanisms at work and we will focus on the advantages resulting from the ability to monitor the energy window lying between the domain of space-borne detectors and ground-based facilities.}

\FullConference{36th International Cosmic Ray Conference -ICRC2019-\\
		July 24th - August 1st, 2019\\
		Madison, WI, U.S.A.}

\begin{document}
\newcommand{\de}{\mathrm{d}\,}

\section{Introduction}
The study of the Universe through the combined investigation of radiation, cosmic rays and gravitational waves, is a major achievement, comparable to the revolutionary astrophysical discoveries that were obtained in the $20^{\rm th}$ century with observations of the sky out of the visible range. After the first detection of gravitational waves from a binary black hole merger \cite{Abbott16}, which inaugurated Gravitational-Wave Astronomy with LIGO/VIRGO, the execution of regular observing campaigns led to the identification of a neutron star merger, with a corresponding electromagnetic counterpart \cite{Abbott17a, Abbott17b}. The event was first detected as a Gamma-Ray Burst (GRB) by the Fermi-GBM instrument, in agreement with the prediction that neutron star mergers are the engine of the family of short GRBs. It subsequently appeared at lower energies as a \emph{kilonova}, providing further information on the site and the evolution of the source \cite{Soares17, Cowperthwaite17, Blanchard17}. On the other hand, the IceCube neutrino experiment announced the first detection of a high-energy neutrino, coming from a direction consistent with the location of TXS 0506+056, a $\gamma$-ray blazar that Fermi-LAT simultaneously detected in a flaring state \cite{IceCube18}, strongly suggesting that jets from Active Galactic Nuclei (AGN) are likely associated with the acceleration of the highest energy particles in cosmic rays.

In both cases, $\gamma$-ray observations were fundamental to obtain prompt information on the electromagnetic counterparts of these first multi-messenger triggers and to address their subsequent identification. Gravitational wave sources associated with neutron stars and the acceleration of relativistic particles in jets are, indeed, expected to produce high-energy photons (HE, $E \geq 10\,$GeV)in close connection with the main event, while low energy electromagnetic signatures mainly arise in the form of late time afterglows. Although our ability to observe the sky in the HE domain has greatly improved over the past decade, we are still missing a wide Field of View (FoV) facility, capable of surveying large fractions of the sky from a few hundred GeV up to several TeV photon energies in the Southern hemisphere. This spectral window is critical, because it corresponds to the energy range which we expect to be covered by the photo-production mechanisms that operate in combination with relativistic particle accelerators and gravitational wave sources \cite{Sironi14}. It is also a fundamental energy range to understand the opacity of the Universe to $\gamma$-ray propagation \cite{Desai19}, therefore leading to important constraints on the amount of background radiation connected with the distribution of light sources, as well as with the possible existence of exotic particles.

Recently, it has been proposed that Extensive Air Shower (EAS) observatories may achieve the large effective areas needed to explore the intrinsically low fluxes of cosmological Very High-Energy sources (VHE, $E \geq 100\,$GeV), in combination with the wide FoVs and high duty cycles required to observe transient phenomena. These properties are currently under investigation with the HAWC observatory and in the LHAASO project \cite{Abeysekara17, Bai19}. Both these facilities, however, are located in the Northern hemisphere, leaving an important fraction of the sky nearly unexplored. In addition, featuring a design that mainly aims at the sensitivity towards the highest energy, they have relevant issues with the sub-TeV domain. Here, we present a study of the astrophysical concepts that could be effectively addressed with the implementation of a new facility, which will operate in the Southern hemisphere down to $100\,$GeV photon energies. We discuss the problems of this energy range and the advantages descending from their solution in the framework of the new Southern Wide-FoV Gamma-ray Observatory collaboration (SWGO), which joins the efforts of international groups that worked on the definition of the desired instrument performance level.

\section{The sub-TeV challenge}
Since $\gamma$-ray photons cannot be deflected and focused to form an image or a spectrum, their detection relies on indirect techniques. At low energies (up to few MeV), the most effective approach is to analyse the energy and direction of the electrons that recoil after a photon is Compton scattered within the detector, while, for higher energies, $e^\pm$ pair production becomes the dominant interaction mechanism. A $\gamma$-ray detector, therefore, has to reconstruct the incoming photon energy and direction with the information extracted from secondary particles. The best way to look for photons up to approximately $100\,$GeV is to work with space-borne instruments, like the Fermi-LAT \cite{Atwood09}, since this type of radiation does not significantly penetrate the Earth's atmosphere. At energies above $E \simeq 100\,$GeV, however, the fact that the majority of astrophysical sources have a power-law decaying spectrum of type $\de N / \de E \propto E^{-\alpha}$, with the spectral index generally being $\alpha \geq 1.5$, would require extremely large collecting areas to record the signal, which, combined with constraints on the size of the calorimeter, imply serious limitations on the detector's acceptance. Photons with these energies are best observed from the ground, by looking at the Cherenkov radiation emitted by relativistic charged particles, produced in the atmospheric showers that are initiated by the primary photons. These showers can be observed either directly in the atmosphere (with Imaging Atmospheric Cherenkov Telescopes, IACTs), or by means of EAS detector arrays located on the ground.

Although the IACT approach offers better sensitivity and spatial resolution, it is affected by the small FoV, which can be covered with an observation, and by severe constraints on the observing time, limited to clear dark nights. The EAS configuration, on the other hand, provides nearly continuous up time and a large FoV, but poorer energy and angular accuracy. Another problem for EAS facilities is that air showers initiated by $\gamma$-rays with $E \simeq 100\,$GeV can only penetrate the atmosphere down to altitudes of order of $5000\,$m a.s.l. and, therefore, require the detectors to operate in peculiar conditions, available in a small set of observing sites. In addition, both techniques suffer from the competing signal due to cosmic-ray initiated air showers. Cosmic rays have an energy spectrum that roughly follows the form:
\begin{equation}
  \frac{\de N_E}{\de E} = 1.8 \left( \frac{E}{\mathrm{GeV}} \right)^{-2.7}\, \mathrm{GeV^{-1}\, s^{-1}\, sr^{-1}\, cm^{-2}}.
\end{equation}
Thus, assuming a spatial resolution element of $1\,$square degree, for a source with the Crab Nebula spectrum above $150\,$GeV, there is approximately 1 $\gamma$-ray shower for $160$ background cosmic rays. It is, therefore, mandatory to have an optimized event reconstruction and background rejection procedure, in order to access this particular energy range.

It has been proposed that an EAS configuration, including a hybrid detector concept with low energy threshold and a time resolution of order $\Delta t < 2\,$ns, can achieve the required performance \cite{Assis18}. More details on the event reconstruction and the subsequent data quality cuts are discussed in associated contributions. Here, we focus on the opportunities that will become available by adopting a strategy that covers the sub-TeV energy range with the sensitivity illustrated in Fig.~1.

\section{Opportunities for the Southern hemisphere Wide-FoV $\gamma$-ray Observatory}
The VHE sky is populated by a large variety of galactic and extragalactic sources, which are known to emit photons up to several TeVs. Combining the observations carried out by Fermi-LAT, HAWC, MAGIC, H.E.S.S. and VERITAS, we have been able to identify several extended and point-like sources. The most prominent diffuse emission comes from the interaction of cosmic rays with the Milky Way inter-stellar medium (ISM) and it is concentrated close to the Galactic Plane. This type of emission shows significant excesses close to Supernova Remnants (SNR) and Pulsar-Wind Nebulae (PWN), providing a striking evidence for their identification as cosmic-ray acceleration sites. Other diffuse emission contributions have been identified in regions close to the Galactic Center, where the extreme crowding of sources makes it difficult to characterize the various components, and in the so-called Fermi Bubbles, two large regions of $\gamma$-ray emitting plasma protruding from the center of the Galaxy, whose actual spectrum is still matter of investigation and debate \cite{Dobler10, Su10, Su12, Ackermann14}.

\begin{figure}[t]
  \begin{center}
    \includegraphics[width=0.8\textwidth]{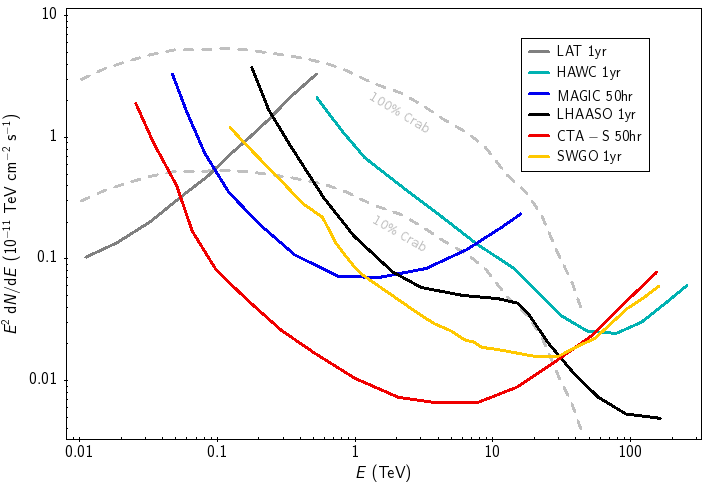}
  \end{center}
  \caption{The proposed sensitivity to a point-like source after 1 year of SWGO observations (orange curve), as compared with Fermi-LAT Pass 8 (gray), HAWC (cyan) and LHAASO (black) sensitivities for the same observing time. The sensitivity with current IACT facilities (MAGIC, blue) and the expected performance of CTA South (red) after 50 hours of observations are also shown for comparison. The feature of the SWGO sensitivity below $600\,$GeV descends from combination of the detector concepts proposed by the LATTES and the SGSO collaborations \cite{Assis18, Albert19}.}
\end{figure}
In the extragalactic domain, on the other hand, the stage is dominated by AGNs, particularly blazars, and GRBs. In both cases, the $\gamma$-ray emission is believed to arise from a relativistic plasma jet and, therefore, to be subject to some extent of beaming. While GRBs can achieve very high luminosities, rivaling for a short time the radiative output of the rest of the Universe, they are transient sources that get disrupted in the process. AGNs, on the contrary, are able to power large scale jets that radiate $\gamma$ rays up to the VHE domain in a more continuous, although significantly variable, way. Fermi-LAT observations of some close AGNs have definitely proven the existence of a diffuse $\gamma$-ray glow around the source, pointing to an emission mechanism that remains active along the structure of the jet \cite{Sun16}.

In most cases, the emission of these extragalactic sources follows a power-law or a log-parabola spectrum, with VHE breaks that may be either intrinsic to the source or due to $\gamma$-ray opacity at high redshift. In the case of AGN flaring activity, however, the presence of excess emission or spectral hardening is often reported. It has also been found that some GRBs exhibit a VHE afterglow, with a few notable cases detected beyond the $100\,$GeV threshold. Although IACTs, and most notably CTA, will certainly provide the best sensitivity to observe sources in this energy range, the short and unpredictable nature of such transients will make their observation particularly hard for instruments with a relatively low duty cycle and the need to be triggered and re-pointed with a certain degree of accuracy.

\begin{figure}[t]
  \begin{center}
    \includegraphics[width=0.45\textwidth]{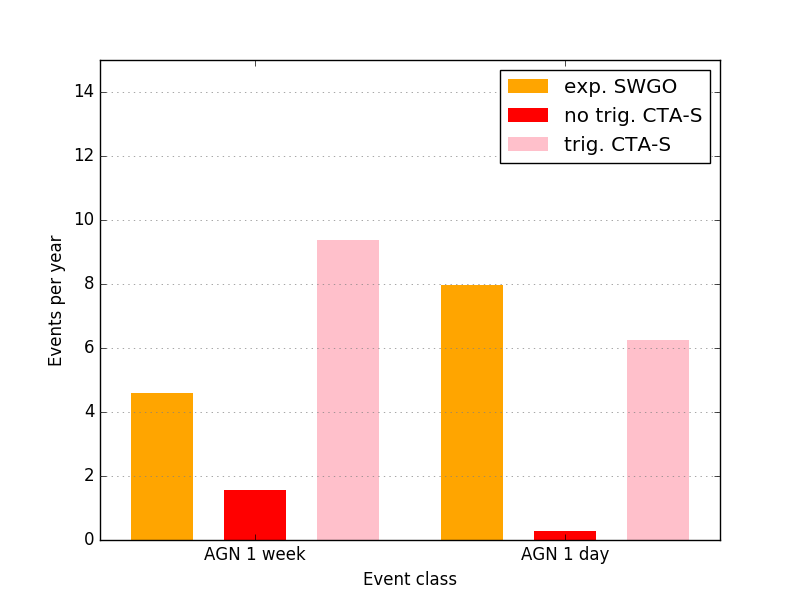}
    \includegraphics[width=0.45\textwidth]{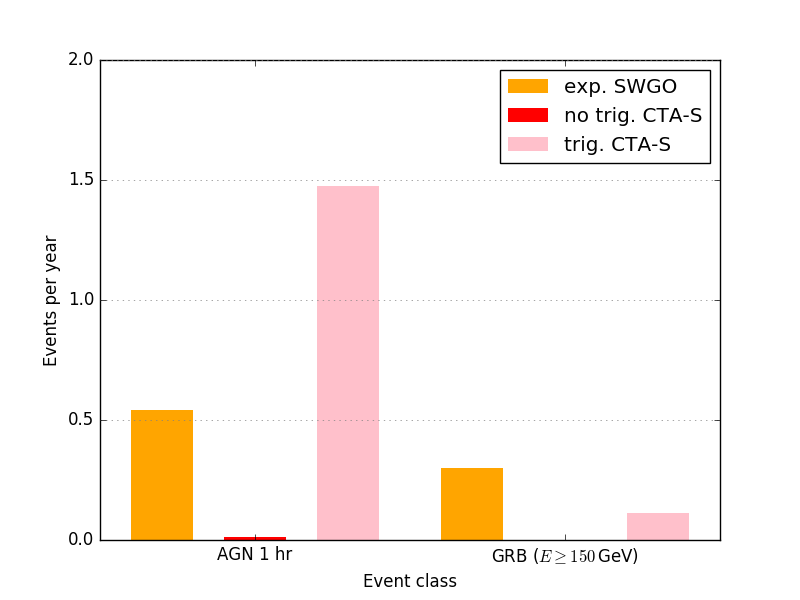}
  \end{center}
  \caption{The estimated rate of long transients (left panel) and short transients (right panel) that are expected to originate photons with $E \geq 150\,$GeV in 1 year. The histograms show the chances to detect different transient types, after accounting for FoV, sensitivity and duty cycle, with SWGO (orange bars), with untriggered CTA observations (red bars) and with CTA alerted by a high energy trigger (pink).}
\end{figure}
An EAS array located in the Southern hemisphere, on the other hand, can provide a relevant contribution to our understanding of the emission processes and the propagation of VHE photons. By defining a target sensitivity curve based on a set of simulated detector concepts, like the one illustrated in Fig.~1, it is possible to predict how this facility compares with the CTA expected performances on different type of sources. If we take the distribution of GRBs and AGN flares detected by Fermi-LAT \cite{Abdollahi17, Ajello19}, it is possible to make predictions on the event detection probabilities of different classes of objects.

Although we currently know little about the VHE emission of transients, it appears quite well established, on the basis of the weekly monitoring carried out in the Fermi-LAT All Sky Variability Analysis (FAVA, \cite{Abdollahi17}), that blazars typically show a spectral hardening, when entering a flaring state. By comparing the distribution of sources and the calculated fluxes with the sensitivities scaled to the available amounts of observing time (in the assumption that short flares are more numerous than long ones), we can estimate the chances of detection of these transients through EAS monitoring or by means of a specifically triggered CTA observation, as shown in Fig.~2. We can follow a similar approach for GRBs, though the unpredictable distribution of the events and the still scarce statistics of VHE detections make this analysis subject to a somewhat higher degree of uncertainty. It has been proven that GRBs can produce VHE photons, but the mechanisms still need to be clarified. Since in 10 years Fermi-LAT has detected 178 GRBs over 2357 GBM triggers, after taking into account the different FoVs and energy ranges and assuming that the number of events producing photons above a minimum energy $E_{low}$ obeys a power-law $N(E > E_{low}) \propto N_0^{-k}$, we can expect that roughly 2-3 events per year are likely radiating above $150\,$GeV. Given that their obsrvation with an IACT requires the trigger to be promptly reported and to occur in favourable observing conditions (clear Moon-less night), the chances of detection for such events would be greatly enhanced by a facility like SWGO, leading to expected detection rates up to 1 event every 1-2 years.

\section{Summary}
We have presented the motivation suggesting that a large FoV $\gamma$-ray observatory should be operated in the Southern hemisphere, with a focus on its sub-TeV capabilities. We discussed the main difficulties that such a facility will have to deal with and we pointed out the motivations to address them. We concluded our discussion with an analysis of the main fields of observation where the instrument is expected to give its major contributions, outlining the scientific questions that it could explore and the way it could operate in combination with the CTA observatory.

Despite the IACT approach remains obviously the best solution for permanent sources and is expected to perform well on long transients (lasting for several days), the EAS contribution becomes relevant and competitive on short transients. A monitoring facility is indeed able to track all the bright events occurring in its FoV, but it can also improve the detection chances of CTA, issuing prompt alerts at energies above a few hundred GeV, where no other experiment is currently effective. On the long time scale, moreover, a wide FoV observatory will produce an extended and well detailed sky map that will certainly provide a fundamental reference to address more advanced studies of interesting regions and serve as a reference to look for variability. With its location in the Southern hemisphere, it will grant an excellent coverage of the Galactic Plane and, particularly, its Center. It will complement the operations of similar experiments running in the Northern hemisphere and it will finally contribute to set up a nearly all-sky prompt alert system for possible multi-messenger triggers.

\end{document}